\documentclass[aps,twocolumn,superscriptaddress,showpacs,nofootinbib]{revtex4}
\usepackage[dvips]{graphics}
\usepackage[dvips]{graphicx}
\usepackage{bm}

\input{epsf}

\newcommand{\be}{\begin{equation}}
\newcommand{\ee}{\end{equation}}
\newcommand{\bea}{\begin{eqnarray}}
\newcommand{\eea}{\end{eqnarray}}

\newcommand{\affA}{Departamento de F\'{\i}sica Te\'orica y del Cosmos,
  Universidad de Granada, Granada-18071, Spain}

\newcommand{\affB}{Laboratoire de Physique Subatomique et Cosmologie,
Universit\'e Joseph Fourier Grenoble 1,
CNRS/IN2P3, \\ Institut  Polytechnique de Grenoble, Grenoble-38026, France }

\begin{document}

\preprint{UGFT-218/07}

\preprint{CAFPE-88/07}

\title{Non-linear metric perturbation enhancement of primordial
  gravitational waves} 
\author{M. Bastero-Gil}
\email{mbg@ugr.es}
\affiliation{\affA}
\affiliation{\affB}

\author{J. Macias-P\'erez}
\email{macias@lpsc.in2p3.fr}

\author{D. Santos}
\email{santos@lpsc.in2p3.fr}
\affiliation{\affB}

\begin{abstract}
We present the evolution of the {\it{ full}} set of Einstein equations 
during preheating after inflation. We study a generic supersymmetric
model of hybrid inflation, integrating fields and metric fluctuations in a
3-dimensional lattice.  We take initial conditions  consistent with
Eintein's constraint equations.  The induced preheating of the  metric
fluctuations is not large enough to backreact onto the fields, but
preheating of the scalar modes does affect the evolution of vector and
tensor modes. In particular, they do enhance the induced stochastic
background of gravitational waves during preheating, giving an energy
density  in general {\it an order of magnitude} larger than that obtained by
evolving the  tensors fluctuations in an homogeneous background metric.  
This enhancement can improve the expectations for detection by planned
gravitational waves observatories.
\medskip                                   

\noindent
keywords: cosmology, inflation, gravitational waves. 
\end{abstract}

\maketitle

Cosmic microwave
background (CMB) measurements \cite{CMB},  are consistent with an
early  period of inflation,  which  
gives rise to the primordial curvature perturbation which
seeds the large scale structure observed today. The observed power
spectrum of temperature anisotropies is
consistent with a Gaussian, adiabatic primordial spectrum, and    
present data sets at most an upper limit on the level of the
primordial tensor contribution predicted by inflation. This
contribution could be 
observed through B-mode polarization by the European Space Agency's Planck 
mission \cite{planck} and future CMB polarization
experiments. Inflation should be followed by a  
reheating period, during which the inflationary vacuum energy is
converted into radiation. During the first stages of reheating, 
 the evolution of the system may be dominated by
nonperturbative effects such as those of preheating,
i.e., parametric amplification of quantum field fluctuations in a
background of oscillating fields \cite{preheating}. Through parametric
resonance, field 
mode amplitudes grow exponentially within certain
resonance bands in $k$ space, this being an efficient way of
transferring vacuum energy into radiation 
\cite{preheating}. It does also enhance the tensor perturbations,
sourced by the field 
anisotropic stress-energy tensor, giving rise to a  
stochastic background of gravitational waves (GWs) \cite{pregw,pregw2,pregw3}.

In most of the studies of preheating of GW,
fields and tensor fluctuations are  evolved in a background 
Friedmann-Robertson-Walker (FRW) metric. However, beyond linear
perturbation, tensors are seeded by the  
other  components of the metric, scalar and vectors    
\cite{secondordergw}. Given that the metric and fields are
nonlinearly coupled through the Einstein equations, the parametric
amplification of field fluctuations will be rapidly transferred to all 
metric perturbations. In a previous paper \cite{pregrenoble} we showed that
this is indeed the case for the scalar metric perturbations. 
We showed that  during the  resonance the scalar metric source term for
the tensors could be comparable to the anisotropic field stress
tensor, arguing that they could  affect the amplification
of tensors.  The purpose of this Letter is to show that this is indeed
the case for hybrid models independently of model parameters. For that
aim we have integrated the full set of Einstein equations, with the
stress-energy tensor provided by the fields in hybrid inflation. 

Einstein equations are written in the so-called 
Baumgarte-Shapiro-Shibata-Nakamura formalism \cite{bssn}, 
widely used because it provides a stable numerical evolution. One
starts with the Arnowitt-Deser-Misner metric, given in terms of  the lapse $N$
and  shift vector $N^i$ functions, and the spatial metric $\gamma_{ij}$:  
\be
ds^2 = N^2 dt^2 - 
\gamma_{ij}(dx^i+ N^i dt) (dx^j+ N^j dt) \,,
\label{metric}
\ee
and performs a conformal transformation of the spatial metric, 
$\tilde \gamma_{ij}= exp(-4 \beta) \gamma_{ij}$, with ${\rm det}
\tilde \gamma_{ij}=1$. The metric dynamical variables are then
$\beta$ and $\tilde \gamma_{ij}$, the trace of the extrinsic
curvature $K= \tilde \gamma^{ij} K_{ij}$, and  its traceless part
$\tilde A_{ij} = exp(-4 \beta) ( K_{ij} - \gamma_{ij}K/3)$. In
addition, a connection variable $\tilde \Gamma^i =
\tilde \Gamma^i_{jk} \tilde \gamma^{jk} = - \partial_j \tilde
\gamma^{ji}$ is introduced in order to compute the Ricci 
curvature more accurately. The lapse and shift vector are gauge
functions, and we choose to work in the synchronous gauge 
and set  $N=1$ and $N^i=0$. 
In this gauge we have $K= 6 \dot \beta$, with  
  $\langle K \rangle/3 =  H(t)$ being 
the average expansion rate, and $\langle e^{2\beta} \rangle= a(t)$  the
average scale factor ( ``$\langle \cdot \rangle$'' denotes spatial
average.). The spatial metric $\tilde \gamma_{ij}$ encodes the two
transverse and traceless degrees of freedom  of the gravity waves,
plus one additional scalar mode and 2 vector degrees of freedom, with 
$\dot {\tilde \gamma}_{ij}= 2 \tilde A_{ij} $.

The matter source is given by the
stress-energy tensor of the two scalar fields (which we take as real
scalar fields) in supersymmetric hybrid inflation, with potential $
V(\phi,\chi)=V_0+ 
g^2 \chi^4/4 + g^2(\Phi^2-\phi_c^2) \chi^2 + m_\phi^2 \Phi^2/2$, where
$\Phi$ is the inflaton field and $\chi$ the waterfall field which 
triggers the phase transition at the end of inflation once the
inflaton  goes below the
critical value $\phi_c$. The vacuum energy $V_0$ is adjusted such that the
potential vanish at the global minimum, $V_0=g^2 \phi_c^4$, and the 
mass for the inflaton field is set by the choice of the slow-roll
parameter $\eta_\phi=m_P^2 m_\phi^2/V_0$, where $m_P$ is the reduced
Planck mass.

The set of equations of motion   are then 
given by the Klein-Gordon equations for the fields in
the curved space given by the metric (\ref{metric}), the equations of
motion for the  connection $\tilde \Gamma^i$, and those   
for the metric variables \cite{bssn} given by
\bea
\ddot \beta + 2 \dot \beta^2 &=& \frac{1}{6m^2_P} ( V - 2 T) -
\frac{1}{24} \dot {\tilde \gamma}_{ij} \dot {\tilde \gamma}^{ij} \,, \label{ddotbeta} \\  
\ddot{\tilde \gamma}_{ij} + K \dot {\tilde \gamma}_{ij} &=&
2 e^{-4 \beta} ( M_{ij}^{TF} - R_{ij}^{TF}) + \dot {\tilde \gamma}^k_j
\dot {\tilde \gamma}_{ki} \,, \label{ddotgamma}
\eea
where $T= (\dot \Phi^2 + \dot \chi^2)/2$ is the
kinetic energy of the fields; the field-dependent source is $M_{ij}=
m_P^{-2} (\partial_i 
\Phi \partial_j \Phi + \partial_i 
\chi \partial_j \chi)$,  $R_{ij}$ is the Ricci tensor of the metric
$\gamma_{ij}$,  and the superscript ``$TF$'' denotes the 
trace-free part of the tensor. 

The system is placed in a finite and discrete 3D comoving box of
length $L$ and $N$ sites per spatial dimension. 
The procedure introduces a comoving
ultraviolet cut-off in both space and momentum. Ideally, one would
like to keep both of them large enough: A big comoving 
box, larger than the initial Hubble size $L > H(0)^{-1}$,  ensures that the
observable initial Universe is within the box; but a large enough
number of points is also required to have all the relevant momentum
modes up to $O(N/L)$.  
For the problem of preheating
after inflation, one tries to optimize the choice of this ratio $O(N/L)$
to have an ultraviolet comoving momentum cut-off still larger than the
preheating cut-off by the end of the simulation. This means that
already at the start of the simulation our comoving box is smaller
than the observable Universe, but the relevant physical modes for
preheating are  all included. 

For the fields  we follow the standard
procedure, with the quantum field theory being replaced by an
equivalent classical field theory. The quantum origin of the fields
remains in the stochastic nature of the initial conditions\footnote{A 
  description of how to follow the quantum-to-classical
  transition for the scalar fields, and how to fix
  the initial conditions for the modes in the classical regime can be
  found in \cite{inifields}. Nevertheless, we have checked that
  for the particular problem at hand, a better determination of the
  initial conditions has little impact on the results.}. 
We start the simulations some
fraction of e-fold $\Delta N_e=0.05$ after the end of inflation, with
the background inflaton field  still close to the critical point, 
$\langle \Phi \rangle= \phi_c {\rm exp}(-\eta_\phi \Delta N_e)$, and
its background velocity given by the slow-roll conditions\footnote{For
  the model parameters considered, 
  the inflaton field reaches the critical point still in the
  slow-regime, and in the absence of any extra mechanism that could
  speed up the inflaton, slow-roll is a good approximation across the
  bifurcation point. For the dependency of the
  GW spectra on the initial velocity of the inflation
  field see \cite{pregw3}.}. The background values for the waterfall field
are set to zero at this point. 
Classical inflaton field fluctuation in a spatial lattice with periodic
boundary conditions are expanded as usual in Fourier modes $\Phi_k$,
with an initial vacuum amplitude  
$|\Phi_k(0)| \simeq 1/\sqrt{2 \omega_k}$,  where $\omega_k=
\sqrt{k^2 + m_\phi^2}$.  
The remaining variables must be chosen to satisfy the Einstein constraint 
equations, the momentum and the Hamiltonian constraint,  at
$t=0$. This is the well known  initial-value problem in general relativity
\cite{initvalueRG}.  We choose initially vanishing tensors and
vectors, i.e., $\tilde \gamma_{ij}(0)=\delta_{ij}$, such that the
momentum constraint reduces to $2 \partial_i K = -
  (3/m_P^2) (\dot \Phi \partial_i \Phi+\dot \chi \partial_i \chi)$. In order to fulfill
this equation, we set the waterfall field  as $\chi(0)= \Phi(0)
-\langle \Phi(0)\rangle$, and\footnote{Given that
  once preheating starts the inflaton and waterfield fluctuations
  become comparable, results are independent of which field we set
  initially in the vacuum, and which is set by the constraint.} 
$\dot \chi(0)= \langle \dot \Phi(0)\rangle
- \dot \Phi(0)$. This allows us to solve the momentum constraint for the
initial value of the expansion rate fluctuations, and use the hamiltonian
constraint  to fix the fluctuations of the scale
factor\footnote{
Violations of the constraints are unavoidable during numerical
integration, and their growth makes the 
system unstable. Numerically, we have checked
that adding  the covariant derivative of the momentum constraint
in Eq. (\ref{ddotbeta}) renders the system more
stable  \cite{constraintevol}.}. 
\begin{figure}[t]
\hfil {\includegraphics[width=8cm,height=6cm]{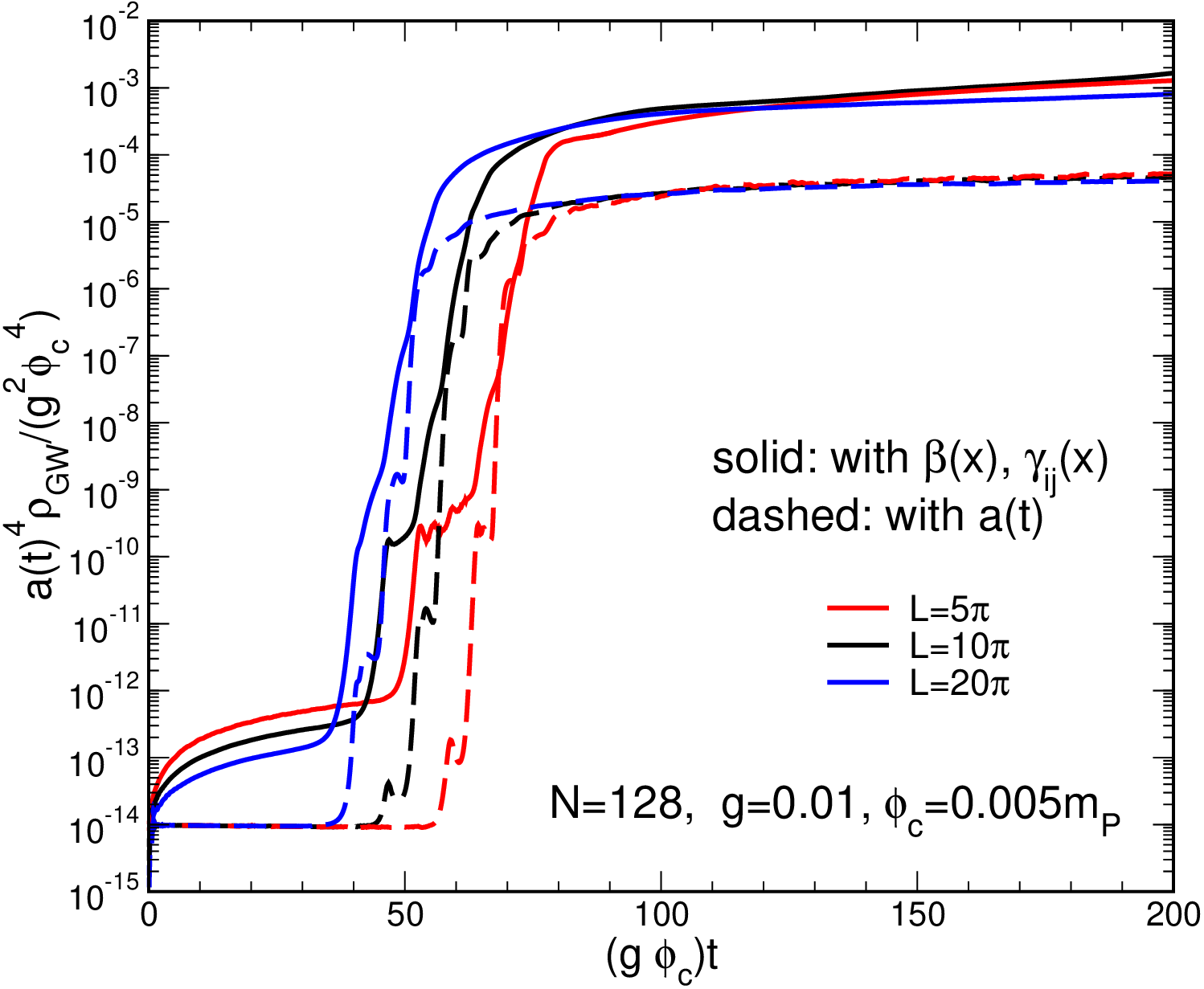}}\hfil
\caption{Evolution of the energy density of GWs, $\rho_{GW}$,
  normalized by the initial vacuum energy $\rho_i$, for different values of the
  box size. Solid lines are the result of integrating the Einstein
  equations; dashed lines show the evolution of $\rho_{GW}$  without
  including other metric fluctuations. 
}
\label{plot1}
\end{figure}

Preheating in hybrid inflation models has been extensively studied in
the literature \cite{hybridpreh}. The
parametric amplification of the fluctuations takes place first through
an spinodal instability for the fields, during which the lower modes
are quickly amplified.   After a few oscillations, the amplitude of
the fields has decayed enough to be out of the spinodal region, and
tachyonic preheating ends. We just follow the evolution of the fields
and metric variables up to the end of the resonance, before we lose
the ultraviolet cut-off for the field modes. Metric variables will follow
 the same pattern of parametric amplification as the fields, with 
 $|\beta|^2 \sim m_P^{-2} |\Phi|^2$. Keeping only the leading $\beta$
 terms in the Ricci tensor, we have: 
\be     
R_{ij}^{TF} 
\simeq [ - 4 \partial_i \beta \partial_j \beta + 2
\partial_i \partial_j \beta ]^{TF} \,, \label{sourceaij}
\ee
and thus $R_{ij}^{TF}$ becomes comparable to the field contribution
$M_{ij}^{TF}$ in Eq. (\ref{ddotgamma}).  

The metric variable $\tilde A_{ij}$ is traceless but nontransverse,
i.e., it contains more degrees of freedom than those two corresponding
to GWs. The traceless and transverse (TT) components are projected by 
using the operator \cite{pregw2} $\Lambda_{ij,lm}(\hat k)=
P_{il}(\hat k) P_{jm}(\hat k) -  P_{ij}(\hat k) P_{lm}(\hat
k)/2$, where $P_{ij}=\delta_{ij}- \hat k_i \hat k_j$ and $\hat
k_i=k_i/k$. The energy density of the GW is then   
$\rho_{GW} = m_P^2 \langle \tilde A_{ij}(t,x)
\tilde A^{ij}(t,x) \rangle^{TT} = m_P^2 \int d^3 k |\tilde
A^{TT}_{ij}(t,k)|^2$ \cite{pregw,pregw2}.  In Fig.~\ref{plot1} we
have plotted 
$\rho_{GW}$, normalized to the initial vacuum energy $\rho_i= g^2
\phi_c^4$. We have taken as parameter models $g=0.01$, 
$\phi_c=0.005 m_P$ and $\eta_\phi=0.05$.  
The value of $\rho_{GW}$ does not depend on the value of the
coupling, which can be rescaled out from all the equations,  but it
does depend on the value of $\phi_c$ \cite{pregw,pregw2,pregw3}, which
sets the scale for the field source term\footnote{
The parameter $\phi_c$ also sets the initial value of the expansion
rate $H_i/(g\phi_c)=\phi_c/(\sqrt{3}m_P)$, and relatively
large values of $H_i$ have an impact on the 
evolution of the resonance, due to the fast redshifting of the field
amplitudes. Tachyonic resonance ends once these amplitudes go below a
certain value, and therefore the larger $\phi_c$ and $H_i$, the shorter
resonance time, and the smaller the amplification of the field
fluctuations and, thus, of GWs.     
} 
such that $\rho_{GW}/\rho_i \propto \phi_c^2$. 
We have included the results for different
choices of the box size, to show that the final value of $\rho_{GW}$ does not
depend on the choice of $L$ as far as we have all the relevant modes
to start and end tachyonic preheating. With a larger comoving box we
have more modes in the low momentum regime, and then tachyonic
resonance starts slightly sooner, as can be seen in the plot. 

\begin{figure}[t]
  {\includegraphics[width=8cm,height=6cm]{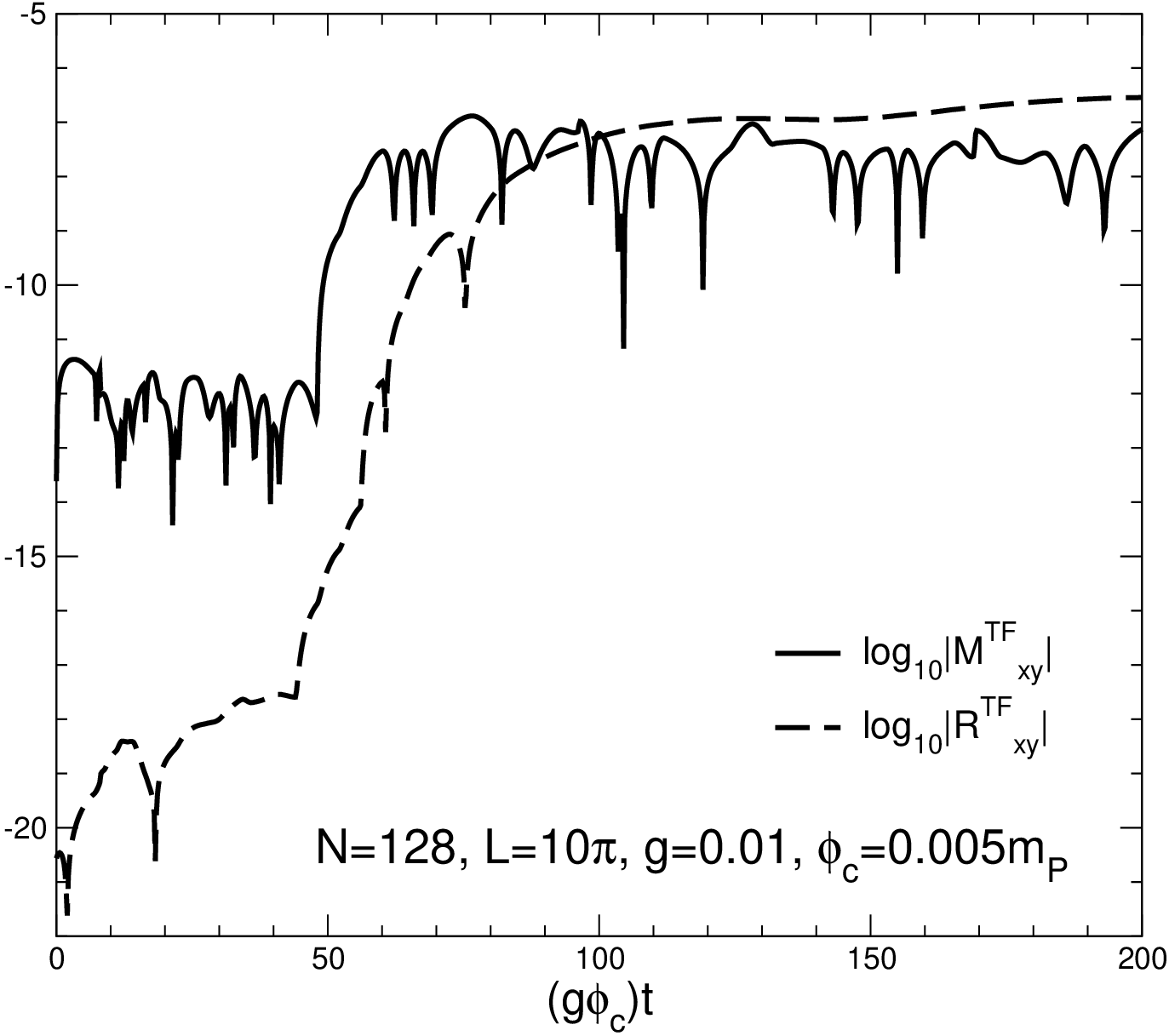}}\hfil
\caption{Comparison of the evolution of the absolute values of the source terms
  $M_{xy}^{TF}$ and $R_{xy}^{TF}$ in Eq. (\ref{sourceaij}), in
  logarithmic scale. The spikes in
  the plot are localized where the corresponding oscillating term vanishes. 
}
\label{plot2}
\end{figure}

We compare in Fig.~\ref{plot1} the results obtained
when integrating the full Einstein equations (solid lines) with those
obtained when integrating the tensor modes in a FRW background
metric (dashed lines). The former are always roughly {\it an order of magnitude
  larger} due mainly to the contribution of the
scalar modes of the metric fluctuations in Eq. (\ref{sourceaij}). 
In Fig.~\ref{plot2}, we show that 
both source terms, $M_{ij}^{TF}$ and the leading term $R_{ij}^{TF}$ in
Eq. (\ref{sourceaij}),  are of the same order by the end
of the resonance. Scalar metric fluctuations start growing immediately after
inflation due to the increase in the kinetic energy of the field [see
Eq. (\ref{ddotbeta})], and this effect leads the initial growth of the
tensor perturbations.  

\begin{figure}[t]
  {\includegraphics[width=8cm,height=6cm]{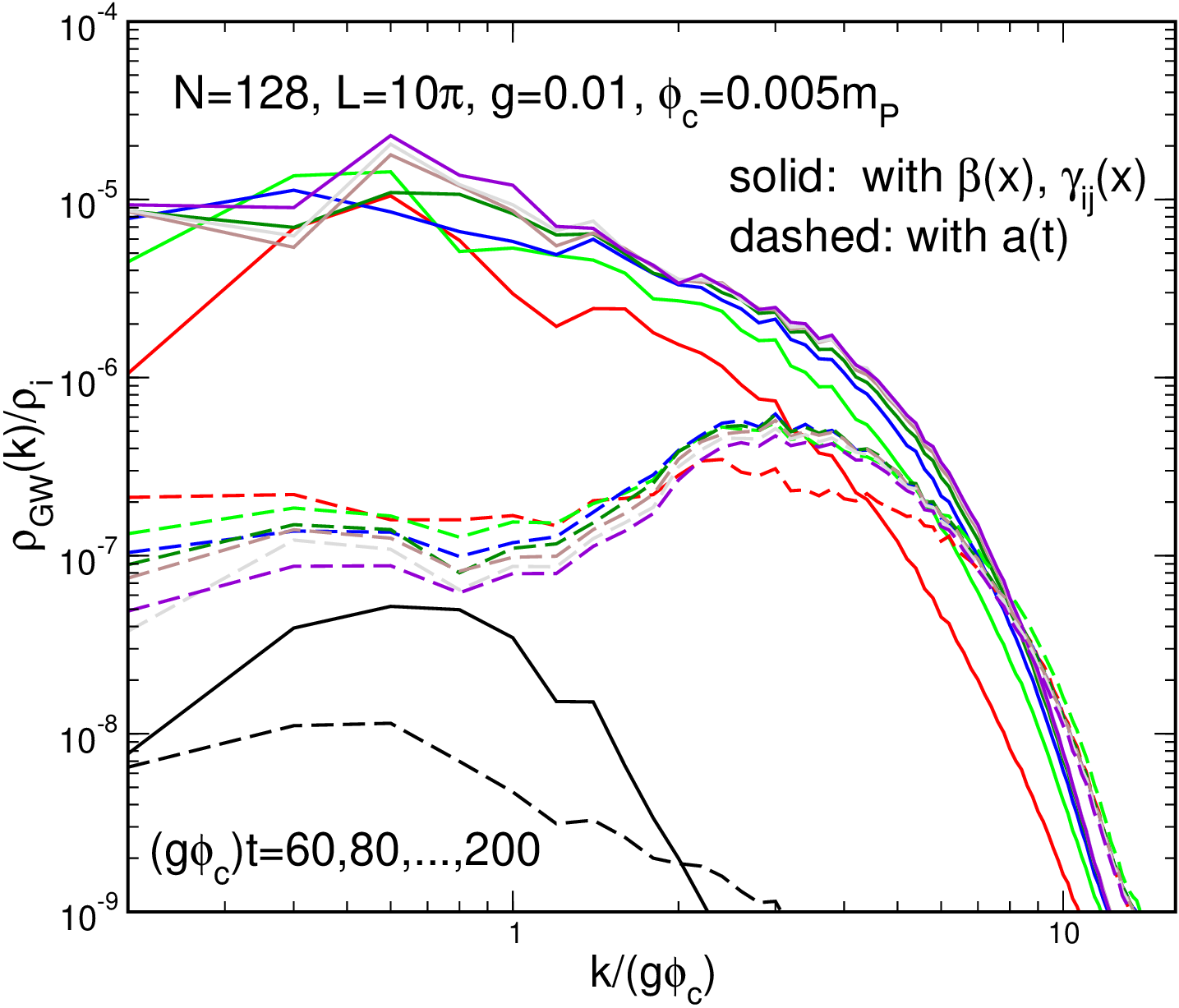}}\hfil
\caption{Spectrum $\rho_{GW}(k)/\rho_i$ per
  logarithmic $k$ interval, for $(g \phi_c) t=60$ to $200$ in
  intervals of $20$ from bottom to top.  
Solid lines are the result of integrating the Einstein equations; 
  dashed lines show the result of the integration in a
  FRW metric.  
}
\label{plot3}
\end{figure}

\begin{figure}[t]
\includegraphics[width=8cm,height=6cm]{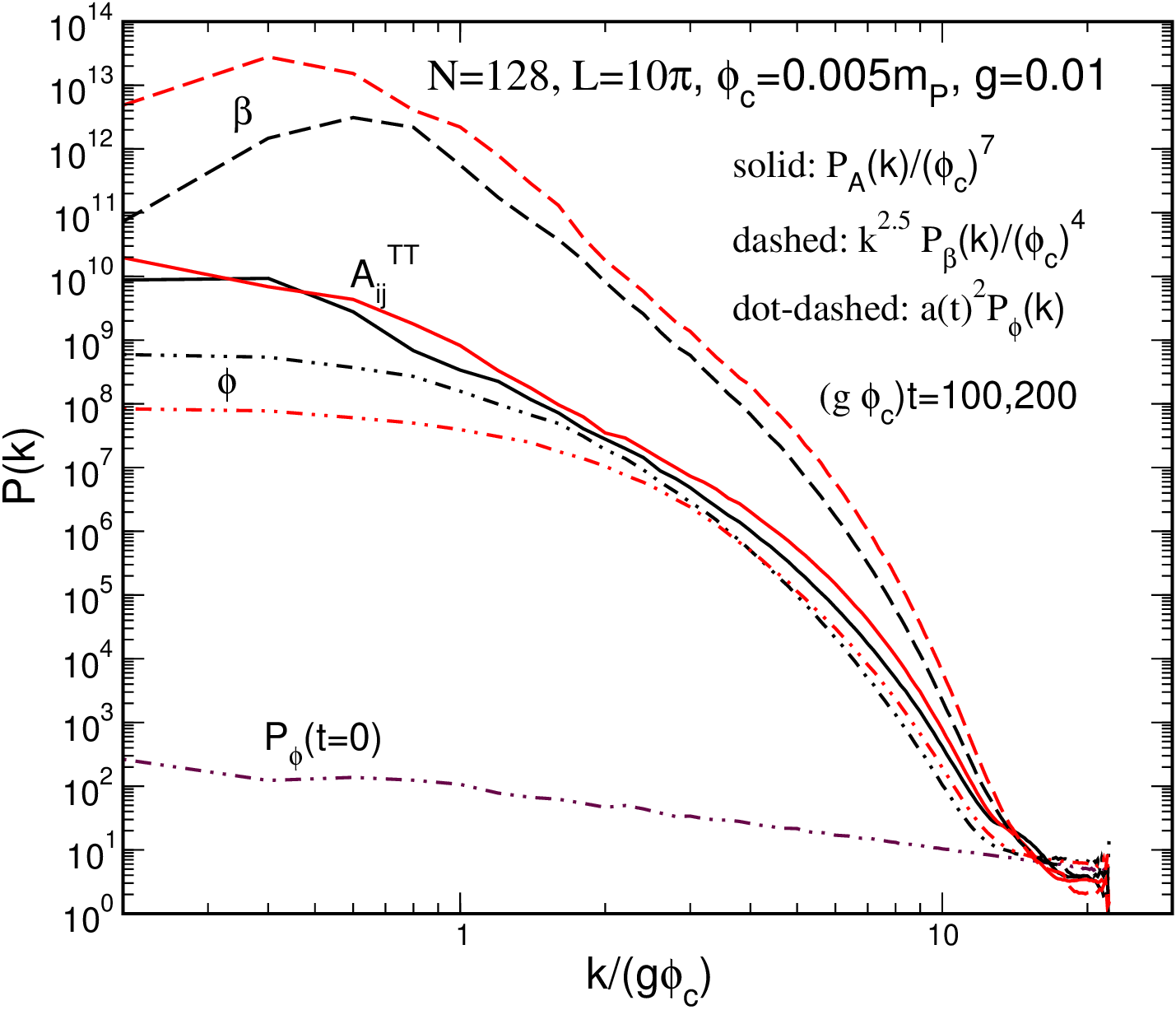}\hfil
\caption{Comparison of
  the power spectrum for the inflaton field $P_\phi(k)$, and metric
  perturbations $P_\beta(k)$  and $P_{A}$, at $(g \phi_c) t=100$ (lower
  curves) and $(g \phi_c)t=200$ (upper curves).   
We have rescaled $P_{A}$ by $\phi_c^{-7}$ to have all
  the spectra in the same scale. For $P_\beta$, we have  
  $k^{2.5} P_{\beta} \simeq \phi_c^4 P_\phi$ from the momentum
  constraint at $t=0$ \cite{pregrenoble}. 
}
\label{plot4}
\end{figure}

In Fig.~\ref{plot3}, we show the spectrum of GW per logarithmic
frequency interval, $\rho_{GW}(k) = d \rho_{GW}/d \ln k$, 
at different times until the end of the tachyonic resonance. Compared with the
spectrum obtained without metric perturbations, the enhancement in the
spectrum is localized at around the peak of the spectrum, shifted
towards lower values at around $k\simeq g\phi_c$. 
This effect at low momenta is again due to the scalar metric
perturbations, which spectrum peaks below  $g \phi_c$. 
This can be seen in Fig.~\ref{plot4}, where  we compare the  power
spectra of the inflaton field fluctuations $P_\phi= |\Phi(k)|^2$, with 
those of the metric fluctuations 
$\beta$ and $A^{TT}_{ij}$, at the end of the
resonance. The field spectrum
at $t=0$ is included as a reference, to check 
that by the end of the simulation the highest momentum modes
have not yet been amplified. 

In summary, we have shown that nonlinear effects due to metric perturbations
enhance the amplitude of the GW stochastic background by {\rm an order of
magnitude} with respect to the calculations in a FRW background
  metric. Taking into account that  
$\rho_{GW}(k)$ is redshifted like radiation after the
resonance, and assuming that entropy is conserved from reheating
onwards,  its  maximum present-day value  normalized by the
critical density today  is given by: 
\be
h^2 \Omega_{GW}^{peak} \simeq 5.5\times 10^{-9} \left(
\frac{\phi_c/m_P}{0.005}
\right)^2 \left(\frac{T_{RH}^4}{\rho_i} \right)^{1/3}\,,
\ee
where $T_{RH}$ is the reheating temperature at which the Universe becomes
radiation-dominated. 
This amplitude is within the reach of the future GW observatory
Advanced-LIGO for $\phi_c \gtrsim 0.005m_P$, or BBO for $\phi_c
\lesssim 0.005m_P$ \cite{GWexper}. 
However, today's values for the frequency are $f= 6.4 \times 10^{10}
\sqrt{g} [k/(g \phi_c)](T_{RH}/\rho_i^{1/4})^{1/3}$ Hz, while the
operating frequency range for Advanced LIGO is $1-10^3$ Hz, and BBO will
operate in the range $10^{-3}-10^2$  Hz.    
Thus, to have the GW spectrum within the observable range, 
the coupling should be at most of the order of $g \simeq
10^{-14}$. 
The numerical simulations so far can  resolve the spectrum for the
typical frequency of the resonance, $k \simeq g \phi_c$ and above, but
not the infrared  tail for subhorizon and superhorizon
modes. The behavior of the GW spectrum in this infrared range and
how they are affected by nonlinear metric effects is still
an open question. A tail of subhorizon modes at the time of preheating
rising slower than $k^3$ could be detected by BBO, although $f_{peak}$
were in the range of $10^{5}$ Hz. 
Similarly, in models where the
primordial curvature is enhanced during preheating \cite{basset}, 
one could expect a similar effect for the tensor primordial spectrum,
up to the level detectable in  CMB polarization experiments. 


We acknowledge for computing resources  the IDRIS-CNRS (Babel)
facility, and the cluster UGRGRID in Granada \cite{computing}. 
We have used the SUNDIALS package, ``SUite of Nonlinear
and Differential, ALgebraic equation Solver'' \cite{sundials} for the
numerical integration.

\end{document}